\documentclass[a4paper]{jpconf}
\usepackage{graphicx}
\begin{document}
\title{Measurement of transverse single spin asymmetries of forward $\eta$ mesons in $p^{\uparrow}+p$ collisions in PHENIX.}

\author{David Kleinjan for the PHENIX collaboration}

\address{University of California Riverside, Riverside, CA 92507}

\ead{david.kleinjan@email.ucr.edu}

\begin{abstract}
The measurement of transverse single spin asymmetries ($A_N$) provides insight into the structure of the nucleon.  Several mechanisms have been proposed that attempt to explain $A_N$ based on QCD, and additional measurements of $A_N$ for different processes further constrain these models.  Using the PHENIX detector at the Relativistic Heavy Ion Collider (RHIC), we study transversely polarized p+p collisions.  Results from PHENIX and other experiments show significant asymmetries in the forward region, which could be due to contributions from both the Sivers and the Collins effects.  Studying the species as well as the kinematic dependencies of these transverse single spin asymmetries will help to disentangle the origin of the observed asymmetries. Therefore, measurements of $A_N$ with inclusive $\eta$ mesons at forward rapidities are an important tool for the understanding of these asymmetries.  In 2008, the PHENIX experiment collected 5.2 pb$^{-1}$ integrated luminosity in $p^\uparrow+p$ collisions at $\sqrt{s}$ = 	200 GeV.  The status of the asymmetry analysis of $\eta$ mesons at forward rapidity will be shown.
\end{abstract} 

\section{Introduction}
\begin{figure}[htbp]
\begin{center}
\includegraphics[width=30pc]{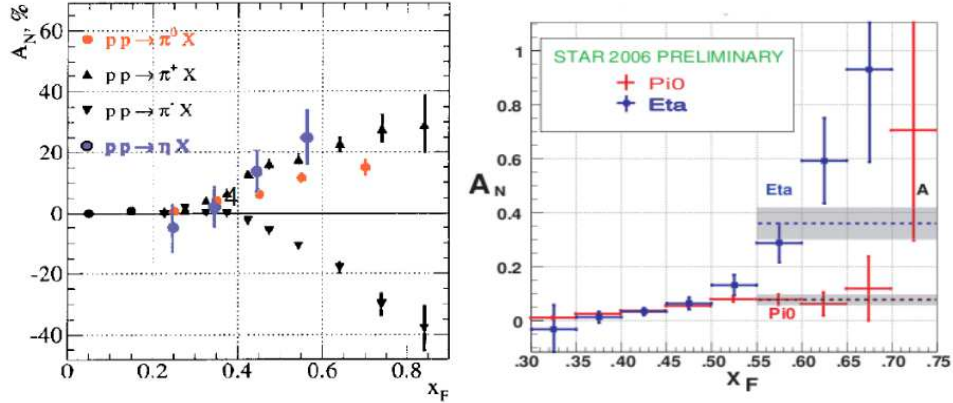}\hspace{1pc}
\end{center}
\caption{\label{fig:etaan}Two previous measurements of $\eta$ meson transverse single spin asymmetry.  The left figure is from FNAL-E704 at $\sqrt{s}$ = 19.4 GeV \cite{bibeta1}.  The right figure is from the STAR experiment at $\sqrt{s}$ = 200 GeV \cite{bibeta2}.}
\end{figure}

Inclusive meson production from transversely polarized $p^\uparrow+p$ collisions have an analyzing power, $A_{N}$, which provides insight into the composite structure of the proton.  The analyzing power, or transverse single spin asymmetry, is quantified  by taking the ratio of the difference and sum of independent polarized processes:

\begin{equation}
\displaystyle A_{N} = \frac{f}{P} \frac{(\sigma^{\uparrow} - \sigma^{\downarrow})}{(\sigma^{\uparrow} + \sigma^{\downarrow})}
\label{eq:an_equ}
\end{equation}
where $P$ is the average beam polarization, $f$ is a geometric scale factor that corrects for the azimuthal acceptance of the detector, and $\sigma^\uparrow$($\sigma^\downarrow$) is the meson production cross section from $p^\uparrow$($p^\downarrow$) + $p$ beam crossings. In prior measurements, non-zero transverse single spin asymmetries have been measured for charged pion production at large Feynman-$x$ ($x_{F}~=~2p_{L}/\sqrt{s}$, where $p_{L}$ is the momentum of the meson along the beam direction) at various energies \cite{bibpion0,bibpion1,bibpion2,bibpion3,bibpion4,bibpion5}.  Measurements of the $\eta$ meson transverse single spin asymmetry have also been made, and are shown in figure \ref{fig:etaan}.  The E704 experiment measured an $\eta$ meson transverse single spin asymmetry comparable with the $\pi^{0}$ transverse single spin asymmetry at $\sqrt{s}$ = 19.4 GeV \cite{bibeta1}, and more recently the STAR experiment measured a very large $\eta$ meson transverse single spin asymmetry at $\sqrt{s}$ = 200 GeV, much larger than the $\pi^{0}$ transverse single spin asymmetry \cite{bibeta2}.

Since collinear pQCD at leading twist predicts a small spin asymmetry due to the hard partonic scattering, initial and final states of the interacting partons and fragmenting hadrons have to be considered.  Two main explanations have been developed to explain these transverse single spin asymmetries, the Sivers and Collins effects. The Sivers effect \cite{sivers} is a correlation between the spin of the proton and the intrinsic transverse momentum of the parton.  The Collins effect \cite{collins} is a correlation between the final state hadron and the initial spin of the proton and parton.

The goal of this analysis is the extraction of the inclusive $\eta$ meson transverse single spin asymmetry at forward rapidity at PHENIX \cite{phenix} for $\sqrt{s}$ = 200 GeV $p^\uparrow + p$ collisions.

\section{Measurement}
For the analysis of the forward $\eta$ meson transverse single spin asymmetry in PHENIX,  the Muon-Piston Calorimeter (MPC) is used.  The MPC consists of two forward electromagnetic calorimeters, referred to as the south (north) MPC, placed $\pm$220 cm from the nominal interaction point, each with a pseudorapidity of 3.1 $<~\eta~<$ 3.9.  Each MPC compromises 196 (220) crystal towers, with dimensions of $2.2 \times 2.2 \times 18~$cm$^{3}$ $PbWO_{4}$.  The MPC is capable of identifying neutral pions  and eta mesons by reconstruction of their decay photons.  PHENIX operates within the RHIC accelerator, which provides two counter-circulating proton beams at various energies up to $\sqrt{s}$ = 500 GeV.  A feature of RHIC is its capability to provide polarized proton beams independently in both directions, allowing for two independent transverse single spin asymmetry measurements by integrating over one beam polarization at a time.

\begin{figure}[t]
\begin{center}
\includegraphics[width=30pc]{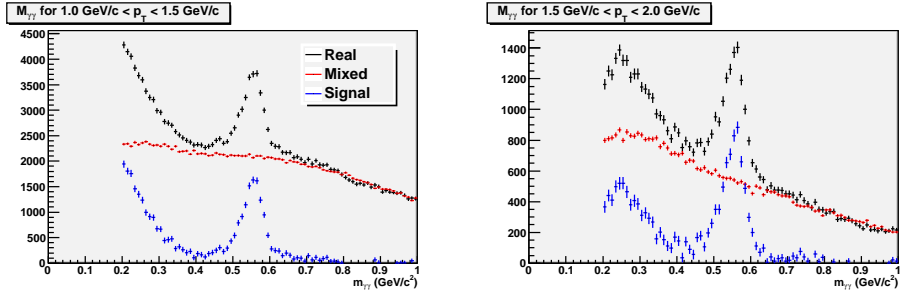}
\end{center}
\caption{\label{fig:minbias}The $\eta$ meson signal extraction from the MB data for two $p_{T}$ bins.  The signal (blue) is extracted by taking the difference between the real event pairs (black) and mixed event pairs (red), which represents the removal of uncorrelated background.}
\end{figure}

Eta mesons are reconstructed and identified in the MPC via the correlation of their decay into two photons.  All pairs of photon candidates are used to form an invariant mass.  Real photon pairs originating from $\eta$ (or $\pi^{0}$) decays will form a pair-mass close to the mass of the $\eta$ (or $\pi^{0}$).  Random combinations (e.g. using a photon from the wrong $\eta$ or $\pi^{0}$ decay or from a mis-identified charged hadron) form a combinatorial background, uncorrelated to the $\eta$ mass.  Figure \ref{fig:minbias} (black curves) illustrates this for two $p_{T}$ bins, where these two effects are indistinguishable on an event-by-event basis.  To remove the combinatorial background, photon candidates are analyzed from different events (necessarily breaking all real combinations) to form a mixed event background (red curves in figure \ref{fig:minbias}). Subtracting real event and mixed event pairs results in a final $\eta$ mass peak which has little background (blue curve in figure \ref{fig:minbias}).  An additional correlated background (in the mass region of 0.2 to 0.4 GeV/$c^{2}$) is observed which is related to a jet background (made up primarily from high energy $\pi^{0} s$).  This is currently being studied in detail using Pythia \cite{PYTHIA} simulations.

\begin{figure}[]%htbp]
\begin{center}
\includegraphics[width=30pc]{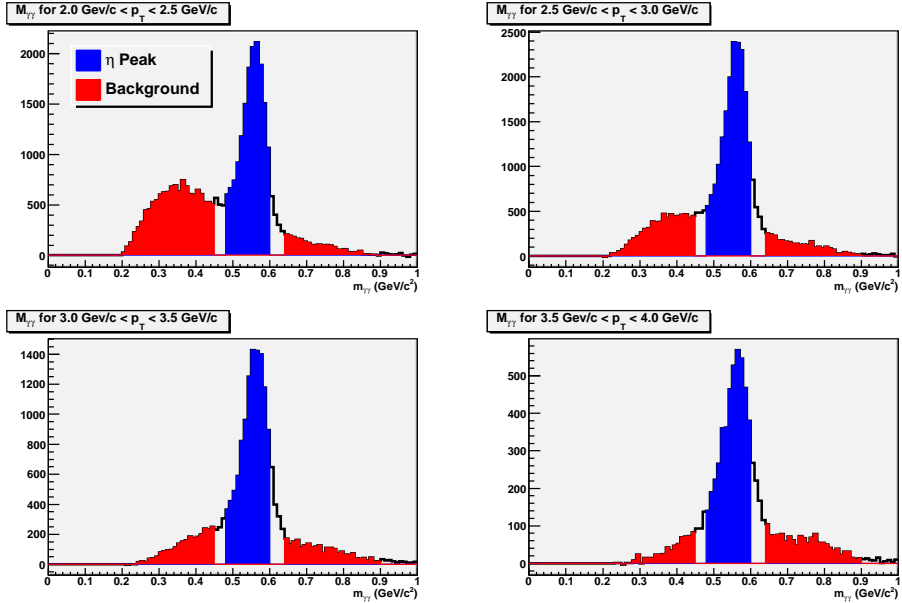}\hspace{1pc}%
\end{center}
\caption{\label{fig:trig}The $\eta$ meson signal extracted from the HEC data for four $p_{T}$ bins. The red bands represent areas where the background asymmetries, $A^{BG}_{N}$, must be calculated and subtracted from the eta peak region, $A^{\eta + BG}_{N}$, to get the  $\eta$ meson $A^{\eta}_{N}$.}
\end{figure}

In order to record the data, two triggers are used: a minimum bias (MB) trigger which uses the Beam-Beam Counters only, and a high energy cluster (HEC) trigger in which a $4 \times 4$ tower energy sum (threshold of $E_{4 \times 4}$ = 20 GeV) in the MPC is used to initiate readout of the detector.  The MB trigger allows for an extraction of the $\eta$ meson yield in the region of 1 $<~p_{T}~<$ 2 GeV/$c$ (see figure \ref{fig:minbias}).  The HEC trigger, accepting only higher energy pairs, reconstruct $\eta$ mesons in the region 2 $<~p_{T}~<$ 4 GeV/$c$ (see figure \ref{fig:trig}).  Similar to the MB data, there is a $\pi^{0}$ dominated jet correlated background that shifts to higher invariant mass as $p_{T}$ increases.

%The MPC records data using two types of triggers:  a minimum bias event (MB) trigger, and a high energy cluster (HEC) trigger, in which a live 4x4 tower energy sum (>20 GeV) in the MPC fires the HEC trigger.  

%The use of the MB trigger allows for an extraction of the $\eta$ meson yield for $\eta$ mesons with a transverse momentum of 1 GeV/$c$ < $p_{T}$ < 2 GeV/$c$.  The Yield extraction for the MB $\eta$ mesons is shown in figure \ref{fig:minbias}.  The signal(blue) is extracted by taking the difference between the real event pairs (black) and mixed events pairs (red).  Real event pairs are formed by making invariant mass pairs from same event collisions, while mixed event pairs are invariant mass pairs made from different event collisions.  Subtracting the mixed event from the real events removes the uncorrelated background.  There is a correlated background at a mass of 0.2-0.4 GeV.  Pythia \cite{PYTHIA} simulations have demonstrated that a majority of this background is from a jet correlated background, made up primarily from high energy $\pi^{0}s$ that merged beyond the resolution of the detector.  This is being studied in more detail.
%The use of the HEC trigger allows an even higher $p_{T}$ $\eta$ meson yield extraction for 2 GeV/$c~<~p_{T}~<~4$ GeV/$c$.  The signal extracted for the MB $\eta$ mesons is shown in figure \ref{fig:trig}.  Similar to the MB data, there is a $\pi^{0}$ jet correlated background that shifts to higher invariant mass as $p_{T}$ increases.

Since the background in both the MB and HEC data is expected to be made primarily from $\pi^{0}s$, the $\eta$ meson $A^{\eta}_{N}$ must be corrected for this background.  This is done by calculating the background asymmetry, $A^{BG}_{N}$, and subtracting it from the $\eta$ peak region asymmetry, $A^{\eta + BG}_{N}$.  The formula used to obtain $A^{\eta}_{N}$ is:

\begin{equation}
\displaystyle A^{\eta}_{N} = \frac{A^{\eta + BG}_{N}-r \cdot {A^{BG}_{N}}}{1-r}, \quad \mathrm {with}  \quad
r\equiv\frac{N^{BG}}{N^{\eta}+N^{BG}}
\label{eq:bk_equ}
\end{equation}
where $r$ is the background fraction under the $\eta$ meson peak, and $A^{\eta + BG}_{N}$, $A^{BG}_{N}$ are calculated from eq. \ref{eq:an_equ}.
% using the yield from the approximate invariant mass region(s) shaded blue(red) in Figure \ref{fig:trig}. 

\begin{figure}[t]
\includegraphics[width=18pc]{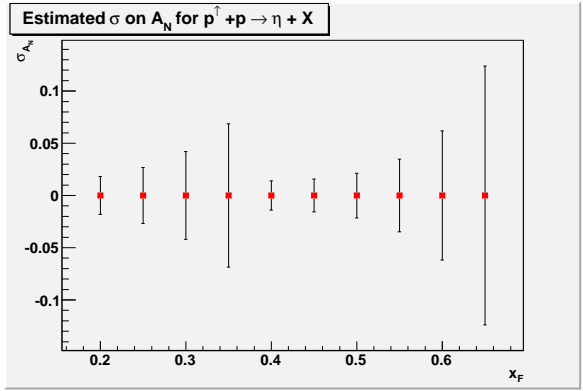}\hspace{2pc}%
\begin{minipage}[b]{18pc}\caption{\label{fig:anerr}Estimated statistical uncertainty on $A_{N}$.  $0.2 < x_{F} < 0.4$ ($0.4 < x_{F} < 0.7$) is calculable from MB(HEC) MPC data.  $x_{F}=0.35$ will provide a consistency check between the MB and HEC data.}
\end{minipage}
\end{figure}

In 2008, PHENIX took 5.2 pb$^{-1}$ of $p^\uparrow + p$ data at $\sqrt{s}$ = 200 GeV, with an average beam polarization of 45\%.  The estimated statistical uncertainty on the $\eta$ meson transverse single spin asymmetry as measured using the MPC is shown in figure \ref{fig:anerr}.

\section{Summary and Outlook}

Using the MPC detector in PHENIX, we have the capability to measure inclusive $\eta$ meson yields at forward rapidity for 1 $<~p_{T}~<$ 4 GeV/$c$ and 0.2 $<~x_{F}~<$ 0.7 from $\sqrt{s}$ = 200 GeV $p^\uparrow + p$ collisions.  The versatility of the RHIC accelerator having both proton beams polarized, as well as having both a north and south MPC in the PHENIX detector, provides an excellent basis for consistency checks on the $\eta$ meson transverse single spin asymmetry measurement, which is expected to be finalized in the near future.


\section*{References}
\begin{thebibliography}{12}

\bibitem{bibpion0} Klem R D et al. (ANL-ZGS) 1976 \emph{Phys. Rev. Lett.} {\bf 36} 929
\bibitem{bibpion1} Antille J et al. (CERN-PS) 1980 \emph{Phys. Lett.}  {\bf B94} 523
\bibitem{bibpion3} Adams D L et al. (FNAL-E581) 1991 \emph{Phys. Lett.}  {\bf B261} 201–206
\bibitem{bibpion4} Adams D L et al. (FNAL-E704) 1991 \emph{Phys. Lett.}  {\bf B264} 462–466
\bibitem{bibpion2} Allgower C E et al. (BNL-AGS)2002 \emph{Phys. Rev.}  {\bf D65} 092008
\bibitem{bibpion5} Arsene I et al. (BRAHMS) 2008 \emph{Phys. Rev. Lett.}  {\bf 101} 042001
\bibitem{bibeta1}  Adams, D L et al. (FNAL-E704) 1998 \emph{Nuc. Phys.}  {\bf B 510} 3-11
\bibitem{bibeta2} Heppelmann, S (STAR) 2009 \emph{Proc.~of XVII Int.~Workshop on Deep-Inelastic Scattering and Related Topics} \verb$http://dx.doi.org/10.3360/dis.2009.195$ 
\bibitem{sivers} Sivers, D W 1990 \emph{Phys. Rev. D}  {\bf 41} 83
\bibitem{collins} Collins, J C 1993 \emph{Nucl. Phys.}  {\bf B396} 161 
\bibitem{phenix} Adcox K et al. 2003 \emph{Nucl. Instrum. Meth.}  {\bf A499} 469–47
\bibitem{PYTHIA} ~Sj\"orstrand T., ~Mrenna S., and ~Skands P. 2006 \emph{JHEP}  {\bf 0605} 026 

\end{thebibliography}
\end{document}